\documentstyle[aps,multicol]{revtex}
\input epsf.sty
\begin{document} 
\draft \title{Energy Gap from Tunneling and Metallic 
Sharvin Contacts onto MgB$_{\mathbf{2}}$:\\ Evidence for a Weakened 
Surface Layer} \author{Herbert Schmidt,$^{1,2}$ J.F.  Zasadzinski,$^{1,2}$ 
K.E.  Gray,$^{1}$ D.G.  Hinks$^{1}$} \address{$^{1}$Materials Sciences 
Division, Argonne National Laboratory, Argonne, IL 60439\\
$^{2}$Physics Division, Illinois Institute of Technology, Chicago, IL 60616} 
\date{February 21, 2001} \maketitle

\begin{abstract}
Point--contact tunnel junctions using a Au tip on sintered $\rm 
MgB_2$ 
pellets reveal a sharp superconducting energy gap that is confirmed by 
subsequent metallic Sharvin contacts made on the same sample.  The 
peak in the tunneling conductance and the Sharvin contact conductance 
follow the BCS form, but the gap values of 4.3 meV are less than the 
weak--coupling BCS value of 5.9 meV for the bulk $T_{\rm c}$ of 39 K.  
The low value of $\Delta$ compared to the BCS value for the bulk 
$T_{\rm c}$ is possibly due to chemical reactions at the surface.
\end{abstract}
\begin{figure}
\centerline{
\begin{minipage}{1\linewidth}
\epsfxsize=1\linewidth \epsfbox{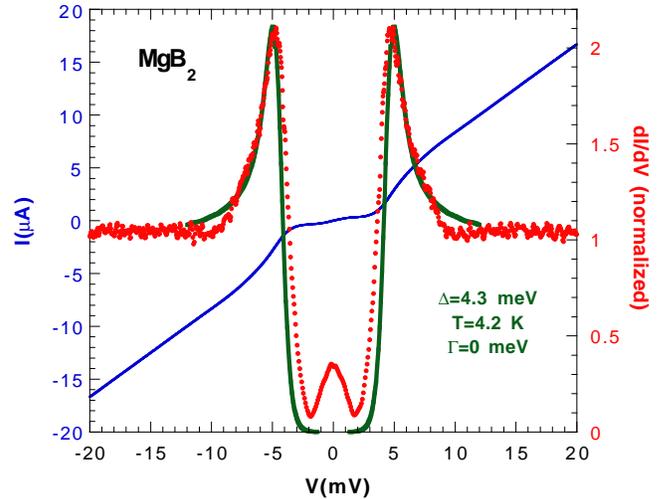} \vskip.16in 
\caption{For a particular point--contact tunnel junction between a Au 
tip and a sintered sample of MgB$_2$, the current (thin line) and the 
differential conductance (small dots) are plotted against voltage $V$ 
at 
a temperature, $T\sim 4.2$ K.  Included is a fit (thick line) to the 
thermally--smeared BCS density of states for $T=4.2$ K.}
\label{fig1}
\end{minipage}}
\end{figure}

\begin{multicols}{2}
The surprising discovery\cite{1} of superconductivity in a known, 
simple 
binary compound, $\rm MgB_2$, with a high $T_{\rm c}$ of 39 K has 
initiated a flurry of activity to understand its properties.  A 
significant isotope effect\cite{2} strongly implies that phonon 
coupling is 
important for the mechanism.  To complete this picture, one should 
observe strong--coupling effects in the electron tunneling density of 
states and connect them via the Eliashberg theory to the recently 
measured\cite{3} phonon density of states.\\ \indent
We report here on 
point--contact tunnel junctions using a Au tip on sintered $\rm MgB_2$ 
pellets that reveal a sharp energy gap, $\Delta$, of 4.3 meV, a value 
that is confirmed by subsequent metallic Sharvin contacts\cite{4} made 
on the same sample.  The peak in the tunneling conductance follows a 
thermally smeared BCS density of states for $T\sim 4.2$ K.  The 
contact resistances varied from $\sim 10$ to 1200 $\Omega$ with the higher 
values showing tunneling characteristics that exhibit a 
superconducting gap at low voltages and at high voltages relatively low 
noise and only a small parabolic conductance correction that implies a 
high--quality tunneling barrier with a height significantly above 100 
meV.  Thus these junctions are attractive candidates to explore 
strong--coupling effects.  Initial STM reports exhibit significantly 
smaller gaps\cite{5} or larger in--gap currents and much greater 
smearing\cite{6}.\\ \indent An additional advantage of point--contact 
tunneling is in its ability to form metallic Sharvin contacts, made {\em in 
situ} using the same sample and Au tip.  The increased conductance 
below $\Delta$ in metallic Sharvin contacts is due to Andreev 
reflections\cite{7}, that are an unmistakable feature of 
superconductivity, whereas well--developed gap structures in 
point--contact tunneling and STM could be due to other effects, like 
charge--density waves\cite{8} or small--particle charging\cite{9}.  We 
find such a conductance increase in our metallic Sharvin contacts that 
is quite sharp, close to the theoretical prediction\cite{7} and gives 
a fit value of $\Delta = 4.3$ meV that is identical to our tunnel 
junctions.  However, in both cases, $\Delta$ is less than the 
weak--coupling BCS value of 5.9 meV for the bulk $T_{\rm c}$ of 39 K, 
and possible explanations are discussed below.\\ \indent The $\rm 
MgB_2$ sample was synthesized from high purity 3 mm diameter Mg rod 
and isotopic $^{11}\rm B$ (Eagle Picher, 98.46 atomic $\%$
$^{11}\rm B$).  The Mg rod was cut into pieces about 4 mm long and 
mixed 
with the $-200$ mesh $^{11}$B powder.  The reaction was done under 
moderate pressure (50 bars) of UHP Argon at 850 $^{\circ}$C.  At this 
temperature the gas--solid reaction was complete in about one hour.  
The sample was contained in a machined BN crucible (Advanced Ceramics 
Corp.  grade HBC) with a close--fitting cover.  There was no reaction 
between the BN crucible and the reactants at the synthesis 
temperature.  X--ray diffraction showed no impurity peaks in the 
powder.  The pellets were made by compacting the synthesized powder in 
a steel die at $\sim 3$ Kb and refiring under the same conditions used 
for the synthesis.\\
\indent Figure \ref{fig1} shows the current, $I$, and the differential 
conductance, ${\rm d}I/{\rm d}V$, plotted against voltage $V$ at a 
temperature, $T\sim 4.2$ K.  Included is a fit to the thermally--
\begin{figure}
\centerline{
\begin{minipage}{1\linewidth}
\vskip-0.21in
\epsfxsize=1\linewidth \epsfbox{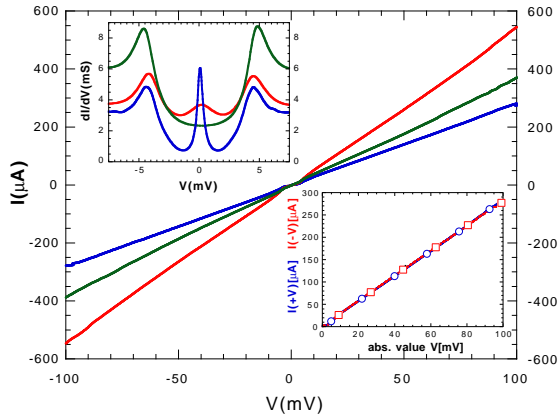} \vskip-0.085in 
\caption{The main panel shows a few more of the large number of 
tunneling junctions made.  The data exhibit only a small parabolic 
term in the high--voltage conductance that implies a high--quality 
tunneling with a barrier height significantly above 100 meV.  Upper 
inset: the low--voltage conductance peaks show a consistent gap of 
order $4-4.5$ meV, but the sub--gap conductivity and zero--bias 
structure varies.  Lower inset: symmetry of the data with respect to 
the voltage--bias polarity is shown by a plot of $|I|$ against $|V|$ 
for both polarities.}
\label{fig2}
\end{minipage}}
\end{figure}
\noindent smeared BCS density of states for $T=4.2$ K, $\Delta=4.3$ meV 
and without a smearing parameter, $\Gamma$. 
Although the agreement in the peak region is acceptable, it is 
possible that slight sample inhomogeneities affect the width and 
height of the conductance peaks.  An excess current is seen at zero 
bias and that can also affect the agreement below the peak voltage.\\ 
\indent Figure \ref{fig2} shows a few more of the large number of 
tunneling junctions made.  Out to voltages of 100 meV, these 
low--resistance point--contact tunnel junctions appear electrically 
stable and show no evidence of heating effects or dielectric 
breakdown: thus the native barrier is suitable for sensitive 
spectroscopy measurements.  The data exhibit only a small parabolic 
term in the high--voltage conductance ($\sim 20$ $\%$ increase at 100
meV) that implies a high--quality tunneling barrier with a height 
significantly above 100 meV.  It is of interest to compare this to 
artificial MgO tunneling barriers that have shown\cite{10} excellent 
tunneling spectra and conductance increases of $\sim 10-15$ $\%$ at 65
meV.  Thus it is possible that the native barrier on our MgB$_2$ 
pellets is MgO.  Strong--coupling effects due to VN phonons ($\sim 1$ 
$\%$ conductance changes) were readily observed in these large area 
(VN--MgO--Pb) junctions\cite{10}.  Thus these point--contact junctions 
on MgB$_2$ are attractive candidates to explore 
strong--coupling effects over the entire phonon spectrum that may 
extend up to $\sim 100$ meV. The upper inset of Fig.  \ref{fig2} shows 
that the conductance peaks at the gap are all of order $4.5-5$ meV, 
but the sub--gap conductivity and zero--bias structure varies.  The 
data are highly symmetric with respect to the voltage--bias polarity, 
as seen in the lower inset to Fig.  \ref{fig2} that plots $|I|$ 
against $|V|$ for both polarities.  This result does not particularly
\begin{figure}
\centerline{
\begin{minipage}{1\linewidth}
\vskip.03in\centerline{\epsfxsize=.875\linewidth 
\epsfbox{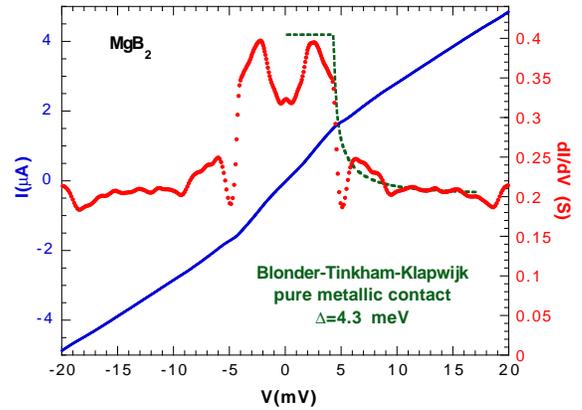}}
\vskip.15in \caption{An equally sharp gap feature (also at 
$\Delta=4.3$ meV) is seen in the metallic Sharvin contacts, that 
exhibit an $I-V$ curve (solid line) and conductance curve (small dots) 
that is quite close to the theoretical prediction \protect\cite{7} 
(dashed line).  The dip at zero bias is similar to that seen 
\protect\cite{14} in metallic Sharvin contacts made on $\rm 
YBa_2Cu_3O_7$ single crystals.}
\label{fig3}
\end{minipage}}
\end{figure}
\noindent support the theory of hole superconductivity\cite{11}.\\
\indent The observation of a conductance peak at zero bias deserves 
further discussion, especially in light of the fact that such peaks 
are used\cite{12} as evidence for a $d$--wave gap symmetry in the 
high--$T_{\rm c}$ cuprates (HTS).  The peaks in HTS junctions arise 
from Andreev bound states that develop when quasiparticles scatter off 
a vacuum interface and change the sign of the order parameter as might 
occur for $d$--wave symmetry\cite{13}.  However, this effect can be 
ruled out as an explanation of our data because of the junction to 
junction variation seen in the upper inset of Fig.  \ref{fig2} and the 
very sharp coherence peaks found at the gap voltage.  The Andreev 
bound state peak removes spectral weight from the coherence peaks and 
thus they should appear considerably broadened as is found 
experimentally\cite{12} in HTS junctions and is predicted 
theoretically\cite{13}.  However, as we have shown, our conductance 
peaks are fit very well by an $s$--wave, BCS density of states, 
broadened by little more than would be expected from thermal smearing 
alone.  Examination of the $I-V$ characteristics in Fig.  \ref{fig1} 
suggests the appearance of a Josephson--like origin for the zero--bias 
conductance peak.  Thus a more likely explanation is a parallel 
conductance channel, perhaps from a filamentary piece of the $\rm 
MgB_{2}$ that has a low--resistance ohmic contact to the Au tip and 
contacts the bulk $\rm MgB_{2}$ with a higher--resistance Josephson 
junction.  When the critical current of this contact is exceeded, its 
large normal--state resistance is shunted by the SIN junction.\\
\indent In addition to the tunneling junctions, manipulation of the Au 
tip could produce the characteristics shown in Fig.  \ref{fig3} of a 
metallic Sharvin contact\cite{4,7}.  Whereas gaps in point--contact 
tunneling and STM could be due to other effects like charge--density 
waves\cite{8} or small--particle charging\cite{9}, the increased conductance 
below $\Delta$ in metallic Sharvin contacts is an 
unmistakable feature of superconductivity, due to Andreev 
reflections\cite{7}. We find an equally sharp gap feature (also at 
$\Delta=4.3$ meV) in the metallic Sharvin contacts, and a conductance 
curve that is quite close to theory\cite{7} (dashed line) that predicts 
a factor of two conductance change.  The dip at zero bias is similar 
to that seen\cite{14} in metallic Sharvin contacts made on $\rm 
YBa_2Cu_3O_7$ single crystals.\\ \indent Since the energy gap value is 
quite robust in a number of junctions and by two techniques, the fact 
that $\Delta$ is consistently smaller than the BCS weak--coupling 
limit must be taken seriously.  In addition, recent NMR 
results\cite{15} imply the bulk gap is $\sim 40$ $\%$
higher than the
weak--coupling
BCS value.  A
possible cause of the small $\Delta$ is a lower $T_{\rm c}$ on the 
surface.  
Weak--coupling BCS theory would require $T_{\rm c}\sim 30$ K in our case and 
about 14 
K for Ref.  \cite{5}. One possibility is a proximity effect with a 
thin layer that is metallic but with a much lower $T_{\rm c}$.  
However, this would be expected to give additional structure  \cite{16} 
in the conductance curve at the bulk gap and our data shows no 
consistent evidence for this. Chemical modifications of the surface 
(e.g., $\rm MgB_2$ reacts with water) can potentially produce a layer 
with a lower $T_{\rm c}$ and a BCS--like density--of--states.  
If the layer is thick 
enough the proximity effect becomes secondary.  A possible scenario is 
linked to the high tunneling barrier that is reminiscent of materials 
like MgO. Since Mg dopes the B layers, any loss of Mg in the 
near--surface regions, to form a MgO cap layer, could lead to lower 
$T_{\rm c}$ and thus $\Delta$.  Hydrogen, or OH$^{-}$, must be 
implicated at least as a catysist, e.g., H$^{+}$ could enter the 
compound interstially or as a replacement for missing Mg$^{2+}$ in the 
$\rm MgB_2$ structure.  It should be pointed out that this would be 
strictly a surface effect, since Mg vacancies are not stable defects 
in the bulk\cite{17}.  Chemically modified surfaces could also explain 
the sample--dependent variations of $\Delta$ (4.3 meV here and 2.0 meV 
in Ref.  \cite{5}).\\ \indent In summary, a well--defined relatively sharp energy gap 
feature at 4.3 meV is seen consistently in metallic Sharvin contacts 
and high--quality (high--barrier) tunnel junctions.  Each are 
consistent with a BCS density of states that is at most only very 
slightly smeared.  The low value of $\Delta$ compared to the bulk 
$T_{\rm c}$ is possibly due to a chemically--modified surface layer 
that could also weaken the intergrain coupling of sintered 
samples\cite{18}.  In order to investigate the intrinsic phonon 
coupling in bulk MgB$_2$ by tunneling, this weakened surface layer 
must be eliminated.  Our observation of a robust native barrier on 
MgB$_2$ could have positive implications for devices, especially if 
thin films become available.\\ \indent
This research is supported by the U.S.  Department of 
Energy, Basic Energy Sciences---Materials Sciences, under contract \# 
W--31--109--ENG--38.

\end{multicols}
\end{document}